\documentclass[lettersize,journal,twoside]{IEEEtran}
\usepackage{amsmath,amsfonts}
\usepackage{algorithmic}
\usepackage{algorithm}
\usepackage{array}
\usepackage{textcomp}
\usepackage{stfloats}
\usepackage{url}
\usepackage{verbatim}
\usepackage{graphicx}
\usepackage{cite}
\usepackage{multirow}
\usepackage{subfigure}
\begin{document}

\title{An Anomaly Detection System Based on Generative Classifiers for Controller Area Network}

\author{Chunheng Zhao, Stefano Longari, Michele Carminati, and Pierluigi Pisu
\thanks{Preprint. Under review.}%
\thanks{Corresponding author: Stefano Longari.}%
\thanks{Chunheng Zhao and Pierluigi Pisu are with the Department of Automotive Engineering, Clemson University, Greenville, SC 29607 USA (e-mail: chunhez@clesmon.edu; pisup@clemson.edu).}%
\thanks{Stefano Longari and Michele Carminati are with the Dipartimento di Elettronica, Informazione e Bioingegneria, Politecnico di Milano, 20133 Milan, Italy (e-mail: stefano.longari@polimi.it; michele.carminati@polimi.it)}}



\maketitle

\begin{abstract}
As electronic systems become increasingly complex and prevalent in modern vehicles, securing onboard networks is crucial, particularly as many of these systems are safety-critical. Researchers have demonstrated that modern vehicles are susceptible to various types of attacks, enabling attackers to gain control and compromise safety-critical electronic systems. Consequently, several Intrusion Detection Systems (IDSs) have been proposed in the literature to detect such cyber-attacks on vehicles. This paper introduces a novel generative classifier-based Intrusion Detection System (IDS) designed for anomaly detection in automotive networks, specifically focusing on the Controller Area Network (CAN). Leveraging variational Bayes, our proposed IDS utilizes a deep latent variable model to construct a causal graph for conditional probabilities. An auto-encoder architecture is utilized to build the classifier to estimate conditional probabilities, which contribute to the final prediction probabilities through Bayesian inference. Comparative evaluations against state-of-the-art IDSs on a public Car-hacking dataset highlight our proposed classifier's superior performance in improving detection accuracy and F1-score. The proposed IDS demonstrates its efficacy by outperforming existing models with limited training data, providing enhanced security assurance for automotive systems.
\end{abstract}


\section{Introduction}
\label{sec:introduction}
The automotive industry has experienced a rapid integration of electronics across various vehicle systems, encompassing functions ranging from entertainment to autonomous driving technologies. Central to these systems are Electronic Control Units (ECUs) interconnected through vehicular networks, predominantly utilizing the Controller Area Network (CAN) \cite{nolte2005automotive}, which is considered
as the de-facto standard in automotive onboard networks \cite{specification2004v2}. This increasing complexity introduces security vulnerabilities, as first introduced by Koscher and Checkoway \cite{checkoway2011comprehensive,koscher2010experimental}, who showcased the potential for remote control by attackers. Since
Miller and Valasek’s demonstration on a Jeep Cherokee in
2014 \cite{miller2014survey}, both industry and academia have raised concerns regarding vehicle security \cite{valasek2014adventures,miller2015remote}. The security weaknesses in CAN systems are well-studied in existing literature, prompting the development of countermeasures and Intrusion Detection Systems (IDSs) in the automotive domain, drawing upon established knowledge from the cybersecurity field. IDSs actively monitor the events within a computer system or network, searching for indicators of unauthorized access. Broadly, in the automotive context, countermeasures and IDSs can be implemented either onboard the vehicle or off-board, covering the entire vehicular ad-hoc network (VANET) or fleet \cite{garg2019sec,garg2020multi,yang2019tree,liang2019mbid}. In this paper, we define an intrusion as a malicious agent that has already gained access to the CAN bus. Thus, our focus is on onboard detection, aiming to prevent attackers from executing exploits leveraging their CAN bus access.

In this paper, we propose a novel intrusion detection model based on a deep generative classifier to identify anomalies. We begin by constructing a causal graph that models the relationships among inputs, outputs, and latent variables. Then, variational Bayes is employed to formulate final prediction probabilities with conditional probabilities derived from the causal graph. A variational auto-encoder (VAE) is utilized to estimate conditional probabilities in Bayesian inference, ultimately leading to the estimation of the final prediction probability. Our experimental evaluation on a publicly available dataset containing attack messages demonstrates the superior performance of our approach, compared to state-of-the-art detection methods.

Our contributions in this work can be summarized as follows:
\begin{itemize}
\item Development of a generative classifier-based intrusion detection model for anomaly detection in CAN traffic by modeling input data distribution.
\item Evaluation of the proposed detection method on a public dataset, benchmarked against state-of-the-art works using multiple classification metrics.
\item The proposed intrusion detection model requires less training data while maintaining competitive performance.
\end{itemize}

The rest of the paper is organized as follows: Section~\ref{sec:related work} summarizes the recent work in the area of IDSs for automotive onboard networks. Section~\ref{sec:methodology} outlines the latent variable model and detection model for the attack classification. In Section~\ref{sec:experiments}, we present experimental results validating the our proposed model on the Car-hacking dataset. Finally, conclusions and discussions are presented in Section~\ref{sec:conclusion}.

\IEEEpubidadjcol

\section{Related Work}
\label{sec:related work}
CAN intrusion detection has been the focus of numerous studies over the years. For a comprehensive review of intra-vehicle IDSs, Jo et al. \cite{jo2021survey} offer a detailed categorization into packet-based and hardware-based systems. Packet-based IDSs can be further divided into flow-based, payload-based, and combined categories. Flow-based IDSs (e.g., \cite{lampe2022ids,seo2018gids,song2016intrusion,taylor2015frequency}) monitor the CAN bus, extracting features such as message frequency or packet inter-arrival time to detect anomalies attempting to overwrite periodic messages. Conversely, payload-based IDSs (e.g., \cite{amato2021can,hanselmann2020canet,kang2016intrusion,longari2020cannolo}) examine the content of CAN packets, typically effective against masquerade attacks. Combined IDSs (e.g., \cite{marchetti2016evaluation,zhang2018two}) integrate both flow-based and payload-based techniques. While flow-based detection achieves high effectiveness against attacks increasing message volume on the CAN bus, with detection rates close to 100\%, it has limitations. Some CAN messages are non-periodic, making it inherently challenging for these systems to detect attacks. Additionally, assuming that the attacker has the capability of silencing the
victim \cite{palanca2017stealth,cho2016error}, or simulate the exact arrival time of
the silenced victim \cite{sagong2018cloaking}, makes it even harder to detect it. Therefore, payload-based and combined IDSs can offer stronger detection by considering specific payload patterns.

Various machine learning techniques have been applied to payload-based and combined CAN IDSs, including from Deep Neural Networks (DNNs) to Generative Adversarial Networks (GANs). Kang and Kang \cite{kang2016intrusion} present a supervised payload-based IDS based on DNN, using mode (i.e., the command state of an ECU) and value information (i.e., the value of the mode) from input features. Several techniques leverage Convolutional Neural Networks (CNNs). For instance, Rec-CNN \cite{desta2022rec} transforms the detection into an image recognition task, utilizing CNNs' image recognition capabilities and generating recurrence plots representing packet time series graphically. Reduced Inception-ResNet \cite{song2020vehicle} simplifies the deep convolutional neural network model of the Inception-ResNet architecture, originally designed for image recognition, to reduce computational time. CANTransfer \cite{tariq2020cantransfer} applies a supervised convolutional LSTM-based model to CAN intrusion detection, aiming for transfer learning to simplify training across different vehicles and IDs. While CNNs are better at processing spatial data, RNNs and LSTMs are generally better suited for temporal data. CAN-ADF \cite{tariq2020can} utilizes employs supervised RNNs and a rule-based IDS to detect simpler attacks, not just identifying them but also classifying them. In contrast, TSP \cite{qin2021application} studies the differences between various loss functions in the development of an LSTM-based IDS. HyDL-IDS \cite{lo2022hybrid} combines CNNs and LSTMs to construct a supervised system that combines both temporal and spatial features of each packet. Autoencoders have also been proposed alongside these techniques to predict or reconstruct time series. O-DAE \cite{lin2020evolutionary} employs a supervised DAE autoencoder to remove noise from the time series of packets and then identifies attacks through the reconstruction. CANet \cite{hanselmann2020canet} is one of the few IDSs that simultaneously elaborates on multiple IDs, theoretically enabling the exploitation of correlations between different information shared through various IDs. It achieves this through an LSTM network per CAN ID and an autoencoder that receives the concatenated output of the various LSTMs. CANnolo \cite{longari2020cannolo} is an unsupervised IDS based on an LSTM autoencoder. A packet window is fed to the RNN autoencoder, attempting to reconstruct it following the trained model. Lastly, the capabilities of GANs have been assessed, with E-GAN \cite{xie2021threat} utilizing an unsupervised GAN for packet layout comprehension and anomaly recognition.

In total, classification-based detection methods can be categorized into two types: discriminative classifiers and generative classifiers \cite{ng2001discriminative}. The previously mentioned detection approaches primarily fall under discriminative classifiers. These classifiers achieve notable success by defining decision boundaries between distinct classes and identifying the most critical features for distinguishing between them. In contrast to discriminative classifiers, generative classifiers aim to model the distribution of each class. This involves understanding how a specific class generates the input data. However, the use of generative classifiers in IDSs has not been extensively explored in the existing literature. In this paper, we contribute to the exploration of generative classifiers in IDSs by developing a model based on variational Bayes.

\section{Methodology}
\label{sec:methodology}
\begin{figure}
  \centering
  \includegraphics[width=1\linewidth]{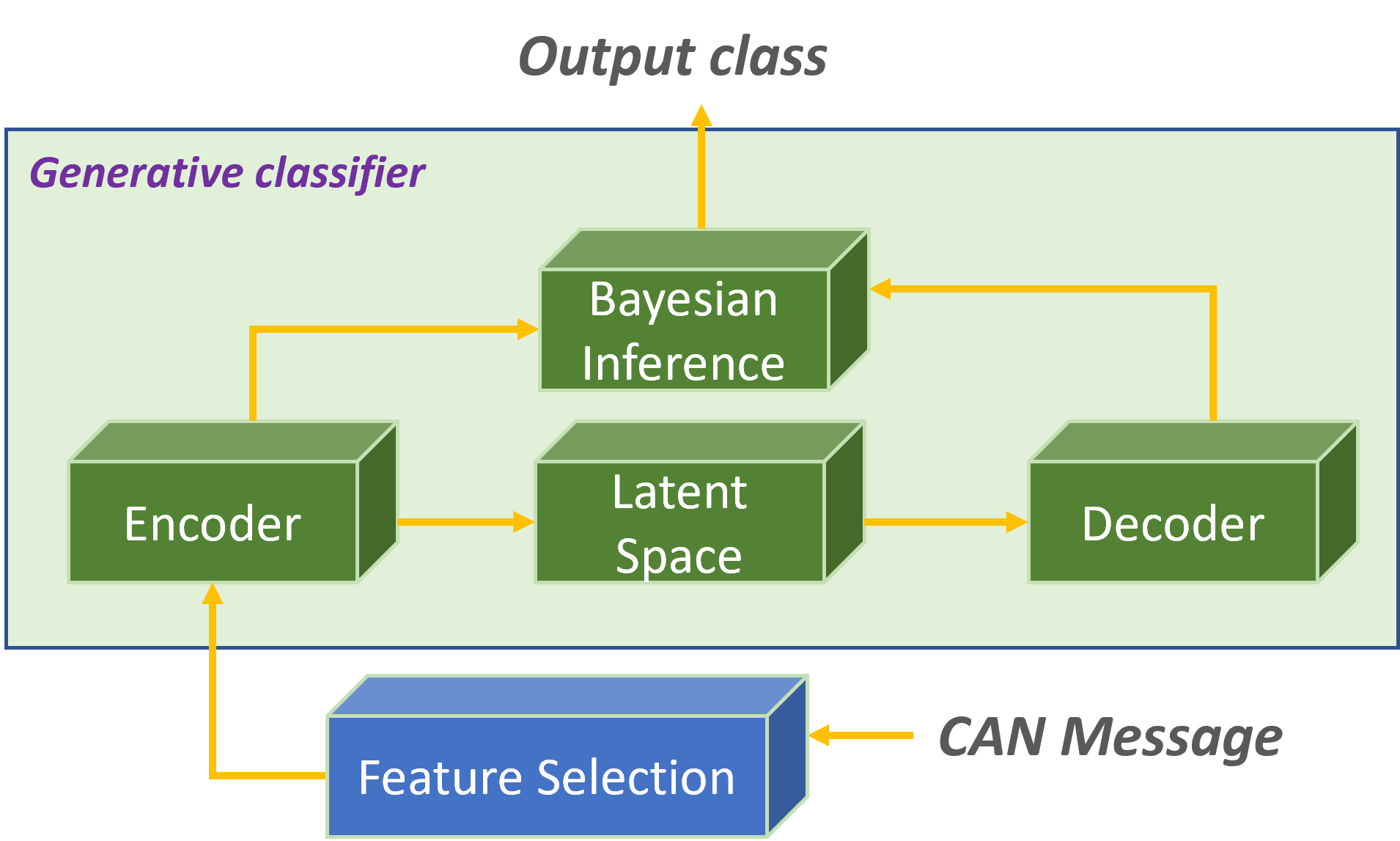}
  \caption{Generative Classifier Architecture with a Variational Auto-encoder.}
  \label{arch}
\end{figure}

In this study, we present a deep generative classifier combined with a deep latent model for detecting malicious messages in the CAN bus. Unlike widely-used discriminative classifiers, generative classifiers aim to model the inherent data distribution of individual classes, specifically delineating how each class produces its input data. Leveraging this characteristic, we could use much less training data to achieve a comparable performance (refer to Section \ref{data} for more details on training and testing dataset size).  Figure~\ref{arch} illustrates the overall architecture of the proposed detector. Raw CAN packets undergo initial processing through a feature selection module to extract relevant features for input into the generative classifiers. Subsequently, a deep latent model is constructed to represent the relationships between inputs, outputs, and latent variables. Finally, a variational auto-encoder conducts the classification process by leveraging Bayesian inference with the causal graph.

\subsection{Threat Modeling}
The first step in building a security system is to understand its potential threats. In this paper, we assume that attackers have already gained access to the CAN bus, enabling them to execute exploits leveraging their CAN bus access. We consider three types of attacks typically found in automotive systems:

\begin{itemize}
\item DoS attack: This attack aims to inject high-priority CAN messages in a short cycle, impairing network availability by occupying nearly all communication resources and causing delays or blocking messages from other nodes.
\item Fuzzy attack: This attack involves injecting fake messages with entirely random data from malicious ECUs, aiming to cause vehicle malfunction.
\item Spoofing attack: This attack aims to deceive the original ECU by spoofing legitimate authentication credentials.
\end{itemize}

Please refer to Subsection~\ref{data} for more details on the attack dataset we use in this paper.

\subsection{Data Preprocessing}
The first module of the IDS is the data preprocessing which extracts relevant features from CAN packets. Standard CAN packets contain multiple data fields (e.g., CAN ID, time stamp, data length, payload and cyclic redundancy check (CRC)), the majority of which are out of the scope of this paper.  Among all the fields, CAN ID, time interval and payload are selected as input features. CAN ID and time interval provide saliency features to detect the DoS attack as this attack typically operate at high frequency with a single ID. Payload and time interval provide relevant features for detecting fuzzy attacks, as fuzzy attacks inject fake packets with random data. Payload and ID can be used to detect spoofing attacks targeting specific ECUs with unique IDs.

The resulting input features are composed by an array $1$ x $k$, where $k$ is the total number of features as shown in (\ref{xfeature}). 
\begin{equation}
X = [X_{id}, X_{t}, X_{p}]
\label{xfeature}
\end{equation}
where $X_{id}$ denotes the CAN ID of the current packet; $X_{t}$ represents the time interval between the current packet and the previous packet with the same ID, which can be calculated using the time stamp; $X_{p}$ represents the payload (i.e., data values) of the current packet. 

\subsection{Deep Latent Variable Model}
\label{modeling}
\begin{figure}[h!]
    \centering
    \includegraphics[width=1\linewidth]{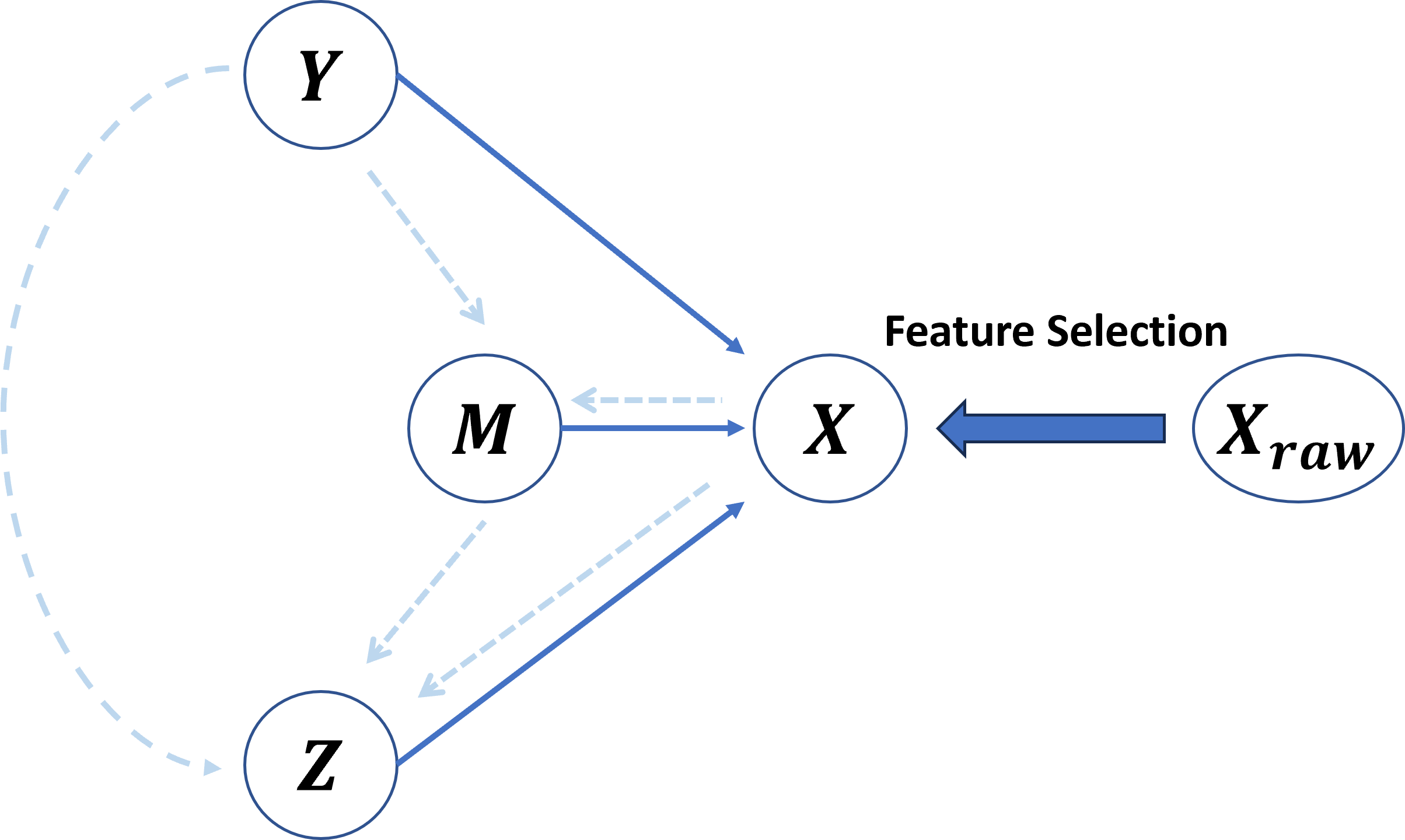}
    \caption{Causal graph. $Y$ is the predicted label, $X$ is the input features, $M$ represents the unseen perturbations and $Z$ represents the rest of latent causes. Solid lines represent the causal reasoning of input data.}
    \label{causal graph}
\end{figure}

To build the generative classifier with latent variables, a causal graph should be created first to model the relationship among inputs, outputs, and latent variables, as shown in Figure~\ref{causal graph}. By leveraging causal reasoning, DNNs can be trained to learn causal relations rather than solely statistical relations between inputs and outputs, thereby avoiding overfitting and improving robustness. In this paper, we define the inputs to the generative classifier as $X$, representing the CAN packet features as described in (\ref{xfeature}). Given these CAN features, multiple factors or causes influence the formation of the input data. Among these factors, $Y$ denotes the label containing the class of attacks, $M$ is a set of variables that can be artificially changed or modified, and $Z$ represents all the other factors that cannot be changed.

Following the structure of the causal graph, we consider variable $M$ as a specific type of perturbation on the input $X$. The perturbation $M$ is generated with respect to the input data $X$ and label $Y$. In this context, $M$ denotes any unknown or unseen attacks. Given that we are utilizing supervised learning which includes all attack labels, none of those attacks are unknown or unseen in the training dataset. Therefore, we set $M$ to 0 during the training phase. However, during the testing phase, $M$ is not constrained to 0 but rather inferred by the network. This approach also enables the model to capture any disparities or biases between the training and testing datasets. Please refer to Subsection~\ref{vae} for more details. Thus for the input data formation, malicious inputs are caused by labels $Y$, unknown attacks $M$ and other latent factors $Z$. 

The causal model representing the forming mechanism of input data can then be formulated as follows:
\begin{equation}
X^{(norm,att)} = P(M,X^{norm},Y,Z)
\label{xadv}
\end{equation}
where $P$ is the process of data formation, $X^{norm}$ is the normal input and $X^{att}$ is the attacked input.

Throughout the training phase, the generative model learns the causal relationships from the input data and subsequently makes accurate classifications based on its reasoning drawn from these factors during the inference phase.

\subsection{VAE-based Generative Classifier}
\label{vae}
After identifying the causal relations between inputs, outputs, and latent variables, we could leverage \eqref{xadv} and variational Bayes \cite{kingma2013auto} to build the generative classifier. Given $Y$ is the target label and $X$ is the input, Bayes' theorem can be applied to estimate the probability of $y$ given $x$ by $p(y|x)=p(y)p(x|y)/p(x)$. In this paper, the generative classifier predicts the label $y$ of an input $x$ as:
\begin{equation}
p(y|x)=\frac{p(x|y)p(y)}{p(x)}=softmax_{c=1}^{C}[\log p(x,y_c)]
\label{eq1}
\end{equation}
where $C$ is the total number of classes and likelihood function $\log p(x,y_c)$ is maximized during the training. During the prediction, the log-likelihood for each $y=c$ is computed for the distribution, then applied with softmax for the final prediction. After including latent variable $m$ and $z$, (\ref{eq1}) can be reformulated as:
\begin{equation}
p(y|x)=softmax_{c=1}^{C}[\log \int p(x,y_c,z,m)\,dm \,dz]
\label{eq2}
\end{equation}

The probability $p(x,y_c,z,m)$ can then be reformulated with latent variables by
\begin{equation}
p(x,y,z,m) = p(x|y,z,m)p(y,z,m)
\label{eq3}
\end{equation}

From the generative modeling process in Figure~\ref{causal graph} (i.e., solid lines), we can then represent $p(y,z,m)$ as:
\begin{equation}
p(y,z,m)=p(m)p(z)p(y)
\label{eq4}
\end{equation}

After substituting $p(x,y,z,m)$ in (\ref{eq2}) with (\ref{eq3}) and (\ref{eq4}), the prediction probability (\ref{eq2}) can be reformulated as:
\begin{equation}
\begin{split}
&p(y|x)=softmax_{c=1}^{C} \\
&[\log \int p(x|y,z,m)p(m)p(z)p(y)\,dm \,dz]
\end{split}
\label{eq5}
\end{equation}

To handle the intractability of the marginal log-likelihood due to the complexity of true posterior $p(z|\cdot)$ and $p(m|\cdot)$ for latent variables with conditional distributions, an approximate distribution \cite{kingma2013auto} $q(z,m;\lambda)$ could be used to approximate the true posterior with variational parameters $\lambda$. Then the model training of maximizing log-likelihood function in (\ref{eq5}) is equivalent to minimizing the divergence between the variational distribution and true distribution. However, this divergence is almost impossible to minimize to zero because the variational distribution is usually not sufficient enough to catch the complexity of the true posterior due to insufficient parameters. To address the issue, Evidence Lower Bound (ELBO) can be adapted here, which is a lower bound on the log marginal probability of the data. Zhang et al. \cite{zhang2018advances} showed that minimizing the divergence is equivalent to maximizing ELBO. ELBO of the log-likelihood in (\ref{eq5}) can be derived using variational posterior $q(z,m;\lambda)$ and Jensen’s inequality as follows:
\begin{equation}
\begin{split}
&\log p(x,y) = \log \int p(x|y,z,m)p(m)p(z)p(y)\,dm \,dz \\
&=\log \int p(x|y,z,m)p(m)p(z)p(y)\frac{q(z,m;\lambda)}{q(z,m;\lambda)} \,dm \,dz \\
&=\log \mathbb{E}_{q(z,m;\lambda)}\left[\frac{p(x|y,z,m)p(m)p(z)p(y)}{q(z,m;\lambda)}\right] \\
&\ge \mathbb{E}_{q(z,m;\lambda)}\left[\log \frac{p(x|y,z,m)p(m)p(z)p(y)}{q(z,m;\lambda)}\right]
\end{split}
\label{eq6}
\end{equation}

Now we can design the inference network (i.e., variational posterior) according to (\ref{eq6}) and the causal graph (Figure~\ref{causal graph}) as follows:
\begin{equation}
\begin{split}
q(z,m;\lambda) &= q_{\delta}(z,m|x,y) \\
&= q_{\delta_1}(z|x,y,m)q_{\delta_2}(m|x,y)
\end{split}
\label{eq7}
\end{equation}
Here the variational parameters are $\delta= \{\delta_1,\delta_2\}$, where $\delta_1$ is parameter for encoder network $q_{\delta_1}(z|x,y,m)$, and $\delta_2$ is parameter for encoder network $q_{\delta_2}(m|x,y)$. Similar variational parameters are defined for the decoder network:
\begin{equation}
p_{\theta}(x,y,z,m)=p_{\theta1}(x|y,z,m)p(m)p(z)p(y)
\label{eq8}
\end{equation}
where the variational parameters are $\theta= \{\theta_1\}$ and $\theta_1$ is parameter for decoder network $p_{\theta_1}(x|y,z,m)$. 

\begin{figure}[h!]
    \centering
    \includegraphics[width=1\linewidth]{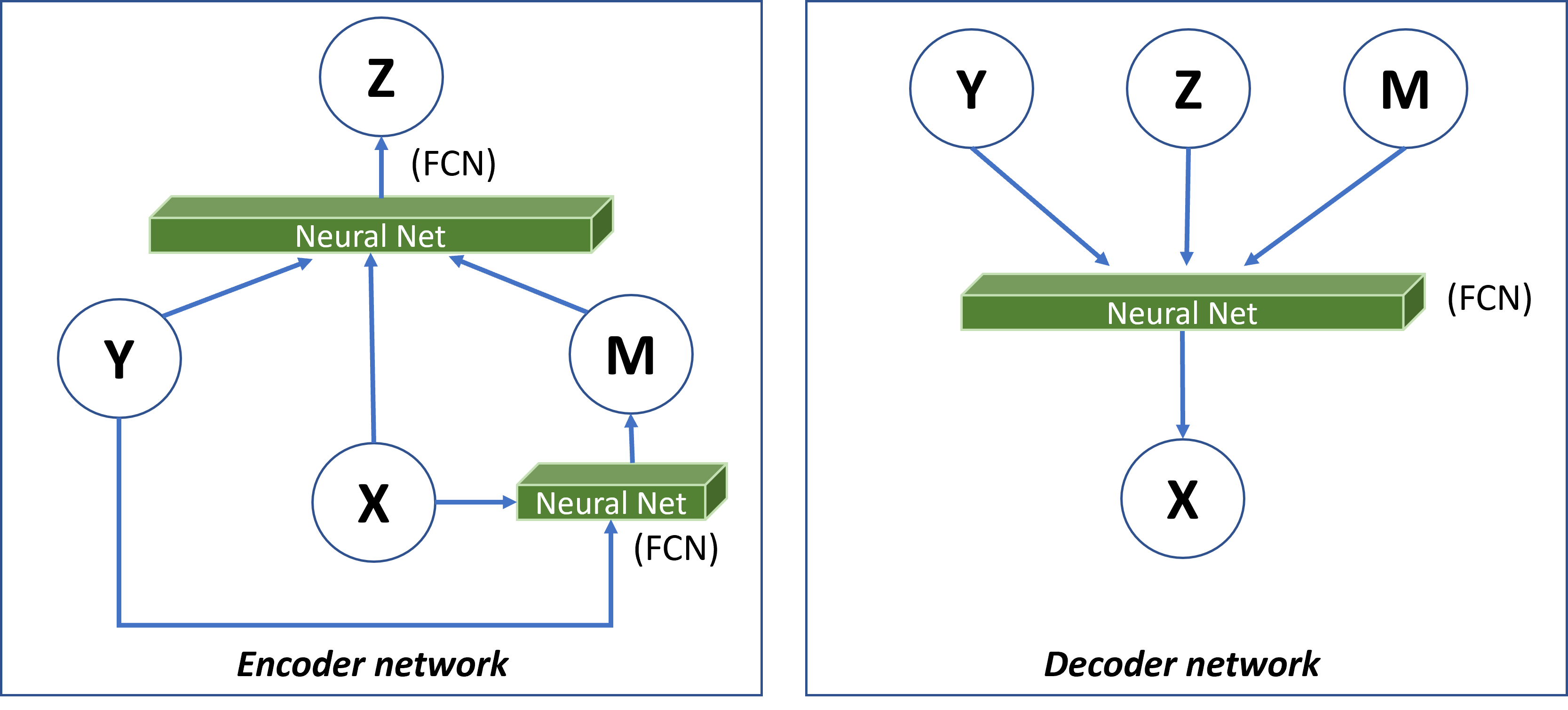}
    \caption{Variational Auto-encoder. Each individual neural net in the encoder and decoder estimates the conditional probabilities for $q$ and $p$, respectively.}
     \label{architecture}
\end{figure}

Figure~\ref{architecture} shows the VAE-based architecture for both the decoder and encoder network containing those separate neural nets. The encoder network is used to compute $q_{\delta}(z,m|x,y)$, where the parameters of CNN is included in $q_{\delta_2}$. The decoder network is used to compute $p_{\theta}(x,y,z,m)$.

After combining (\ref{eq6}), (\ref{eq7}) and (\ref{eq8}), the training of $p_{\theta}$ and $q_{\delta}$ network on a dataset $S$ with $N$ samples can be done by maximizing the lower-bound function:
\begin{equation}
\mathbb{E}_{S}=\sum_{n=1}^N\mathbb{E}_{q_{\delta}}\left[\log \frac{p(z)p(y_c)p_{\theta 1}(x|y,z,m)}{q_{\delta 1}(z|x,y,m)}\right]
\label{train}
\end{equation}

$m$ is set to $0$ during the training time. The prior distribution of $p(z)$ is set with $\mu = 0$ and $\sigma = 0$, where $\mu$ and $\sigma$ represent mean and variance respectively. The prior distribution of $p(y)$ is set according to the total classes in the dataset. During the model inference period, $m_t$ is not set to $0$ but instead sampled from $q_{\delta_2}(m|x,y_c)$ and $z_t$ is sampled from $q_{\delta_1}(z|x,y_c,m_t)$. The prediction probability can be obtained by:
\begin{equation}
\begin{split}
&p(y|x)=\frac{p(x|y)p(y)}{p(x)}\simeq \\ &softmax^C_{c=1} \left[\log \sum_{k=1}^K \frac{p(z)p(y_c)p_{\theta 1}(x|y_c,z_t,m_t)}{q_{\delta 1}(z_t|x,y_c,m_t)}\right]
\end{split}
\label{pred}
\end{equation}
where $C$ is the number of classes, and $K$ is the number of samples.

\begin{table}
\centering
\caption{\textbf{Attack Data Type and Size \cite{song2020vehicle}.}}
\label{table}
\begin{tabular}{|c|c|c|}
\hline
Attack type&
No. of CAN messages&
No. of attack messages\\
\hline
DoS attack &3,665,771 &587,521  \\
\hline
FUZZY attack &3,838,860 &491,847  \\
\hline
GEAR attack &4,443,142 &597,252 \\
\hline
RPM attack &4,621,702 &654,897 \\
\hline
\end{tabular}
\label{tab1}
\end{table}

\section{Experimental Evaluation}
\label{sec:experiments}
\subsection{Datasets}
\label{data}
The Car-hacking dataset \cite{song2020vehicle} is constructed by logging CAN traffic via the onboard diagnostic (OBD-II) port from Hyundai’s YF Sonata while conducting message injection attacks. The dataset comprises four primary data features: timestamp, identifier (ID, in hexadecimal format), data length code (DLC, ranging from 0 to 8), data payload (8 bytes), and the label of a CAN message (i.e., attack or normal). As shown in Table~\ref{tab1}, it contains a total of 14,237,978 normal CAN messages and 2,331,497 anomaly messages, which are categorized into three attack types, resulting in four attacks in total (i.e., DoS attack, fuzzy attack, spoofing the drive gear, and spoofing the RPM gauge). 

\noindent\textbf{DoS attack:} Messages of ‘0000’ CAN ID are injected every 0.3 ms, which is the most dominant ID. As the ECU that attempts to send a message with the most dominant CAN ID always wins the bus, other ECUs are prevented from transmitting their messages. 

\noindent\textbf{Fuzzy attack:} Messages with random CAN IDs and payloads are injected every 0.5 ms.

\noindent\textbf{Spoofing attack:} Messages of certain CAN IDs related to RPM/gear information are injected every 1 ms, changing the RPM gauge and drive gear information on the instrument panel.

Note in the original dataset \cite{song2020vehicle}, each attack type was treated as an independent dataset and not as a multi-class dataset. However, in our paper, as we aim to distinguish between each different class of attacks, we combine and mix all four datasets, creating a multi-class classification dataset. The mixed dataset is then divided into a training dataset and a testing dataset with a ratio of 3:1, containing 12,500,000 and 4,000,000 messages, respectively. Leveraging the inherent characteristics of generative modeling, which effectively models input data distribution, we could train the network with a relatively small subset of data. In this paper, we selected 360,000 CAN messages from the training dataset, allocating 90,000 messages for each of the four attack types. To address the class imbalance in the original dataset, where normal data significantly outnumbers attack data, we fixed a 2:1 ratio of normal to attack messages within each 90,000-message subset. Consequently, the final training dataset contains only 360,000 messages, including 240,000 normal messages and 120,000 attack messages. The testing dataset remains unchanged with 4,000,000 messages. Thus, the ratio of the final training dataset to the testing dataset is approximately 0.09. Despite the small amount of training data, our model achieves remarkable performance, as demonstrated in the Section \ref{results}.

\subsection{VAE Parameters}
\label{model}
The Fully Connected Networks (FCNs) estimating $q_{\delta_1}(z|x,y,m)$ and $q_{\delta_2}(m|x,y)$ both consist of 10 hidden fully-connected layers, each comprising 26 neurons with Rectified Linear Unit (ReLU) activation functions. Additionally, the FCN estimating $p_{\theta_1}(x|y,z,m)$ consists of 10 hidden fully-connected layers, each containing 36 neurons with ReLU activation. or training optimization, we employ the Adam optimizer with a learning rate of $1\mathrm{e}{-4}$, while the batch size was set to 100. The training process involves a total of 250 iterations.

\begin{figure}
  \centering
  \subfigure[Accuracy]{
  \includegraphics[width=0.8\linewidth]{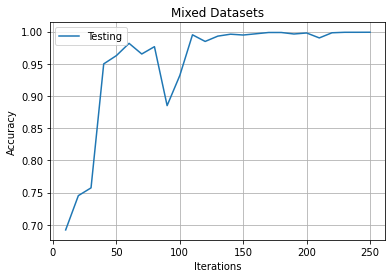}}
  \subfigure[False Positive \& Negative Rate]{
  \includegraphics[width=0.8\linewidth]{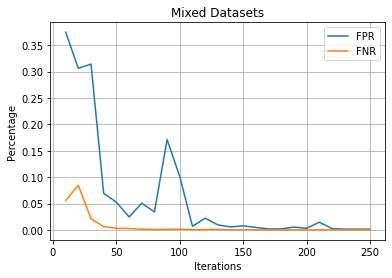}}
  \caption{Overall Accuracy, False Positive Rate (FPR), and False Negative Rate (FNR) vs. Iterations.}
  \label{acc}
\end{figure}

\begin{table}
\centering
\caption{\textbf{Overall Accuracy, False Positive Rate (FPR), and False Negative Rate (FNR).}}
\label{table}
\begin{tabular}{|c|c|c|c|}
\hline
&
Accuracy&
FPR&
FNR\\
\hline
Training &99.98\% &0.05\% &0.00\% \\
\hline
Testing &99.97\% &0.03\% &0.00\% \\
\hline
\end{tabular}
\label{tab1}
\end{table}

\begin{table*}[t!]
\centering
\caption{\textbf{Detection Performance Comparison of State-of-the-art IDSs}}
\label{table}
\begin{tabular}{|c|c|c|c|c|c|c|}
\hline
IDSs&
Attacks&
Accuracy&
Precisions&
TPR&
FPR&
F1-score\\
\hline
\multirow{4}{*}{Reduced Inception-ResNet \cite{song2020vehicle}} & DoS Attack & 0.9993 & 0.9995 & 0.9963 & 0.0001 & 0.9980 \\
&Fuzzy Attack & 0.8730 & 0 & 0 & 0.0002 & - \\
&Gear Spoofing Attack & 0.8223 & 0 & 0 & 0.0001 & - \\
&RPM Spoofing Attack & 0.7774 & 0 & 0 & 0.0003 & - \\
\hline
\multirow{4}{*}{CANTransfer \cite{tariq2020cantransfer}} & DoS Attack & 0.9991 &0.9990 &0.9951 &0.0002 &0.9971 \\
&Fuzzy Attack & 0.8718 &0 &0 &0.0001 &- \\
&Gear Spoofing Attack & 0.8223 &0 &0 &0.0004 &- \\
&RPM Spoofing Attack & 0.7774 &0 &0 &0.0003 &- \\
\hline
\multirow{4}{*}{CAN-ADF \cite{tariq2020can}} & DoS Attack & 0.9938 &0.9826 &0.9785 &0.0033 &0.9805 \\
&Fuzzy Attack & 0.8715 &0.0505 &0.0002 &0.0006 &0.0004 \\
&Gear Spoofing Attack & 0.8222 &0 &0 &0.0004 &- \\
&RPM Spoofing Attack & 0.7769 &0.1200 &0.0005 &0.0012 &0.0011 \\
\hline
\multirow{4}{*}{TSP \cite{qin2021application}} & DoS Attack & 0.9802 &0.9100 &0.9728 &0.0183& 0.9403 \\
&Fuzzy Attack & 0.8714 &0 &0 &0.0005 &- \\
&Gear Spoofing Attack & 0.8221 &0 &0 &0.0005 &- \\
&RPM Spoofing Attack & 0.7774  &0 &0 &0.0003 &-  \\
\hline
\multirow{4}{*}{O-DAE \cite{lin2020evolutionary}} & DoS Attack & 0.9933 &0.9742 &0.9843 &0.0050& 0.9792 \\
&Fuzzy Attack & 0.8714 &0 &0 &0.0006 &- \\
&Gear Spoofing Attack & 0.8222 &0 &0 &0.0004 &- \\
&RPM Spoofing Attack & 0.7774  &0 &0 &0.0003 &-  \\
\hline
\multirow{4}{*}{LDAN \cite{zhao2021efficient}} & DoS Attack & 0.9806 &0.9099 &0.9756 &0.0184& 0.9416 \\
&Fuzzy Attack & 0.8717 &0 &0 &0.0006 &- \\
&Gear Spoofing Attack & 0.8224 &0 &0 &0.0001 &- \\
&RPM Spoofing Attack & 0.7775  &0 &0 &0.0002 &-  \\
\hline
\multirow{4}{*}{E-GAN \cite{xie2021threat}} & DoS Attack & 0.9806 &0.9099 &0.9756 &0.0184& 0.9416 \\
&Fuzzy Attack & 0.8717 &0 &0 &0.0002 &- \\
&Gear Spoofing Attack & 0.8224 &0 &0 &0.0001 &- \\
&RPM Spoofing Attack & 0.7774  &0 &0 &0.0003 &-  \\
\hline
\multirow{4}{*}{HyDL-IDS \cite{lo2022hybrid}} & DoS Attack & 0.9936 &0.9819 &0.9781 &0.0034& 0.9800 \\
&Fuzzy Attack & 0.8715 &0.0612 &0.0002 &0.0005 &0.0005 \\
&Gear Spoofing Attack & 0.8221 &0 &0 &0.0001 &- \\
&RPM Spoofing Attack & 0.7769  &0.1042 &0.0005 &0.0011 &0.0009  \\
\hline
\multirow{4}{*}{CANet \cite{hanselmann2020canet}} & DoS Attack & 0.9993 &0.9992 &0.9966 &0.0014& 0.9979 \\
&Fuzzy Attack & 0.8717 &0 &0 &0.0002 &- \\
&Gear Spoofing Attack & 0.8223 &0 &0 &0.0001 &- \\
&RPM Spoofing Attack & 0.7774  &0 &0 &0.0003 &-  \\
\hline
\multirow{4}{*}{Rec-CNN \cite{desta2022rec}} & DoS Attack & 0.9803 &0.9097 &0.9740 &0.0185& 0.9408 \\
&Fuzzy Attack & 0.8714 &0 &0 &0.0006 &- \\
&Gear Spoofing Attack & 0.8221 &0 &0 &0.0005 &- \\
&RPM Spoofing Attack & 0.7774  &0 &0 &0.0003 &-  \\
\hline
\multirow{4}{*}{As proposed in this paper} & DoS Attack & $\mathbf{0.9999}$ & 0.9749 & 0.9750 & $\mathbf{8.0231e-07}$ & 0.9749 \\
&Fuzzy Attack & $\mathbf{0.9998}$ & 0.9456 & 0.9511 & 0.0002 & 0.9478 \\
&Gear Spoofing Attack & $\mathbf{0.9999}$ & 0.9737 & 0.9754 & $\mathbf{0.0001}$ & 0.9744 \\
&RPM Spoofing Attack & $\mathbf{0.9999}$ & 0.9831 & 0.9832 & $\mathbf{5.3584e-07}$ & 0.9832 \\
\hline
\end{tabular}
\label{tab2}
\end{table*}

\subsection{Results}
\label{results}
Figure~\ref{acc} illustrates the testing accuracy, false positive rate (FPR), and false negative rate (FNR) trends across iterations during training. Table~\ref{tab1} presents the final accuracy, FPR, and FNR achieved after 250 iterations of training. Our model demonstrates remarkable performance, achieving 99.97\% accuracy in multi-class classification on the mixed dataset. Additionally, it achieves a low FPR of around 0.03\% and an FNR of 0.00\%.

To facilitate a comprehensive comparison with other machine learning techniques prevalent in the state of the art, we adopt the experimental evaluation framework outlined by Wang et al. \cite{wang2022analysis} on the public Car-hacking dataset. We follow the same experimental procedure with our proposed generative classifier-based IDS. The performance metrics utilized include accuracy, precision, true positive rate, false positive rate, and F1-score. Table~\ref{tab2} presents the results obtained by Wang et al., followed by ours. For detecting DoS attacks and RPM spoofing attacks, the proposed IDS achieves perfect performance with 99.99\% detection rate and nearly 0\% FPR. Similarly, for detecting the other 2 attacks, we achieve nearly perfect performance with over 99.98\% detection rate and a FPR of less than 0.0002. Overall, our proposed IDS exhibits superior detection rates across all categories of attacks and achieves comparable metrics in cases where it does not outperform existing methods.

\section{Conclusion}
\label{sec:conclusion}
In this paper, we introduce a novel Intrusion Detection System (IDS) based on a generative classifier with a deep latent model, designed specifically for anomaly detection in automotive CAN communications. Our proposed generative classifier utilizes a latent variable model to construct a causal graph and subsequently estimates prediction probabilities using variantional Bayes. The incorporation of a variational auto-encoder in the architecture facilitates the prediction of conditional probabilities. 

For evaluation, we conduct a comprehensive comparison with state-of-the-art Intrusion Detection Systems (IDSs) designed for CAN, utilizing the public Car-hacking dataset with attack messages. Benefiting from the characteristic of generative modeling, which can capture the inherent data distribution of individual classes, we are able to use significantly less data for training. The results demonstrate that our proposed classifier outperforms current methods across various metrics, enhancing detection accuracy. In future work, we aim to test the IDS with a more complex real-world dataset, exploring its robustness and efficacy under diverse conditions. Additionally, we plan to analyze the impact of attacks specifically targeting the deep learning-based IDS itself, such as adversarial attacks. Furthermore, given the incorporation of a causal model into the classifier, we will emphasize investigating the explainability capabilities of our proposed IDS.


\bibliographystyle{IEEEtran}
\bibliography{main}

\end{document}